\documentclass[12pt]{article}
\usepackage{graphicx}
\usepackage{epsf}
\usepackage{amsmath}	% Advanced maths commands
\usepackage{amssymb}	% Extra maths symbols
\makeatletter
\def\@biblabel#1{}
\makeatother

\begin{document}

%\begin{titlepage}

\begin{center}

\noindent
{\bf EXTRAORDINARY SUPERNOVA iPTF14hls: AN ATTEMPT AT INTERPRETATION}

\bigskip

N. N. CHUGAI

\bigskip

{\it Institute of Astronomy of Russian Academy of Sciences, Moscow}\\

\end{center}

\vspace{0.5cm}
email: $<$nchugai@inasan.ru$>$

\vspace{1cm}
\begin{abstract}
It is shown that the H$\alpha$ luminosity and the Thomson optical depth of the 
iPTF14hls 
on day 600 after the detedtion provide us with the esimate of the envelope age  
which turns to be about 1000 days. I propose a model that suggests an explosion 
of a massive star with the radius of $\sim 2\times10^{13}$ cm at 450 days prior 
to the discovery. For the optimal model the ejected mass is $30\,M_{\odot}$, 
and the kinetic energy is $8\times10^{51}$ erg. The energy source at the dominant 
luminosity stage is presumably related to the relativistic bipolar outflow originated 
from a disk accretion onto the black hole. The [O\,I] 6300, 6364 \AA\ doublet in 
the spectrum on day 600 is shown to be the result of the emission of at least 
$1-3\,M_{\odot}$ of oxygen in the ejecta inner zone. The oxygen distribution is 
non-spherical and can be represented either by two components with blue and red shifts 
(in the optically thin case), or by one blueshifted component, in the case of optically 
thick lines for the filling factor of $\sim 2\times10^{-3}$.
\end{abstract}

\section{Introduction}

Recently Arcavi et al. (2017) reported photometric and spectral data on the highly 
unusual supernova iPTF14hls. The phenomenon is distinguished by its high 
luminosity and the extended duration of the powerful luminosity exceeding 600 days.
At first thought the large luminosity duration might originate from the 
circumstellar (CS) interaction. However authors emphasise that the spectrum 
of iPTF14hls is different from those of SN~IIn and therefore the interaction is 
rulled out (Arcavi et al. 2017). Authors suggest that the prolonged luminosity 
is caused by the black hole accretion power because this mechanism permits ones at 
least qualitatively to account for significant luminosity variations which could be 
a problem for the magnetar mechanism. From the H$\alpha$ absorption 
they conclude that the spectrum originates from an envelope of several tens of solar 
masses and kinetic energy of $\sim 10^{52}$ erg ejected a few hundred days before a 
"terminal explosion".

Dessart (2018) proposes a model in which an explosion of the $15\,M_{\odot}$ star  
just prior to the discovery ejects the envelope of $13\,M_{\odot}$ with the 
kinetic energy of $1.3\times10^{51}$ erg. The prolonged luminosity in this model 
is caused by a magnetar. The reported synthetic spectra adequately reproduce 
the evolution of the quasicontinua and the strong H$\alpha$ with the luminosity of
$3\times10^{41}$ erg s$^{-1}$ on day 600 (Dessart 2018). Yet author emphasises difficulties 
of his model: (i) the lack of Na\,I doublet and Ca\,II triplet and (ii) the lack 
of the H$\alpha$ absorption at the high velocities. The first problem stems from the 
high model ionization, which is related to the large luminosity and 
the model parameters. The second problem suggests that the model kinetic energy is insufficient to provide larger density at the high velocities.
At present we thus have two scenarios for the  iPTF14hls phenomenon: 
(i) a massive star explosion with the ejection of the several tens solar mass envelope 
occured several hundreds days before the discovery (Arcavi et al. 2017); 
(ii) an explosion coeval with the discovery of the moderately massive star with the ejection 
of the $13\,M_{\odot}$ enevelope that is ionized and heated by the magnetar (Dessart 2018).

Andrews and Smith (2018) report an interesting spectrum of iPTF14hls on 
day 1153. It shows boxy H$\alpha$ with the luminosity of $4\times10^{39}$ erg s$^{-1}$ that 
is attributed to the belt-like emission region arising from the circumstellar 
interaction. This poses a question on the search for signatures of the CS interaction 
at the earlier phase; this however unlikely is feasible 
given the two orders of magnitude larger H$\alpha$ luminosity on day 600.
Recently Woosley (2018) discussed possible scenarios for iPTF14hls 
including CS interaction, pulsational pair-instability supernovae, magnetar model and 
demostrated difficulties for all the mentioned options. 

The proposed study is aimed at the search for additional arguments in favour 
of a possible scenario for iPTF14hls upon the basis of the data analysis.
On the basis of simple considerations we will argue in favour of large envelope 
mass and the explosion long before the iPTF14hls detection. 
A simple approach will be used to model the bolometric light curve of iPTF14hls 
in the scenario of black hole accretion; the H$\alpha$ and Na\,I modelling 
will be found to be consistent with overall scanario. Interesting data on the oxygen mass, velocities and distribution will be recovered from the 
[O\,I] 6300, 6364 \AA\ doublet in the spectrum on day 600.

Following Arcavi et al. we count time from the discovery moment 22/09/2014 and 
adopt for iPTF14hls the luminosity distance of 156 Mpc.

\section{General considerations}

Archive photometric data before the iPTF14hls detection (Arcavi et al. 2017) admit 
the explosion during 117 day interval prior ro the discovery or in the range 470 - 350 days before 
the detection. Arguments based on a simple model and observational data permit us to 
distiguish between two possibilities. Indeed, adopting homogeneous freely expanding envelope 
it is easy to write the envelope radius expressed in terms of the H$\alpha$ luminosity 
($L_{32}$) and the Thomson optical depth ($\tau_T$)
\begin{equation}
R = \frac{3L_{32}}{4\pi \alpha_{32}E_{23}}\left(\frac{\tau_T}{\sigma_T}\right)^{-2} = 
1.1\times10^{16}(L_{32}/10^{41}\mbox{erg/s})\tau_T^{-2} \quad\mbox{cm}\,,
\end{equation}
where $\alpha_{32} = 3.1\times10^{-13}$ cm$^3$\,s$^{-1}$ is the effective recombination coefficient for 
the H$\alpha$ in the case C (optically thick balmer lines) assuming $T = 5500$\,K,  $E_{23}$ is 
the energy of the H$\alpha$ photon, $\sigma_T$ is the Thomson cross section.
Adopting the H$\alpha$ luminosity on day 600 $L_{32} \approx 3\times10^{41}$ erg s$^{-1}$
(Dessart 2018, Arcavi et al. 2017) and the Thomson optical depth  $\tau_T = 0.8$ recovered from the H$\alpha$ on day 600 (see below) one gets $R \approx 5.3\times10^{16}$ cm. 
With the expansion velocity of $5000$ km s$^{-1}$ estimated from the H$\beta$ absorption 
(Arcavi et al. 2017) we obtain 
the age for the envelope $\approx 1200$ days on day 600. This suggests that the preferred explosion 
epoch is between 400 - 500 days before the discovery. The equation (1) combined with the expression for 
the Thomson optical depth results in the relation between the envelope mass and the 
hydrogen ionization fraction  $M = 17x\,M_{\odot}$. The requirement $x \lesssim 0.5$ 
[high ionization prevents the emergence of Na\,I doublet (Dessart 2018)] suggests the large 
envelope mass, $\gtrsim 30\,M_{\odot}$.

If the prolonged iPTF14hls luminosity is actually caused by the black hole accretion then
one should admit that the exit on the maximal luminosity occured with a significant delay 
after the explosion to avoid a conflict with archive flux upper limits.
Although a mechanism of this delay is unknown, the very fact of flux variability in the light 
curve admits such a delay of accretion, e.g. due to a slow evolution of the disk 
outer region. For a reasonable black hole mass (10-40$M_{\odot}$) the required accretion luminosity 
($\sim 10^{43}$ erg s$^{-1}$) by four orders exceeds the Eddington limit. One should admit, therefore, that 
the power injection into the envelope is fulfilled as the kinetic energy of the relativistic bipolar outflow (RBO); this term may include colimated jets and the less collimated disk wind.

The RBO interacts with the supernova envelope via the reverese shock in which the flow is "thermalized" 
thus forming the bubble filled with relativistic particles and magnetic field. The picture is reminiscent 
of the interaction of a pulsar wind with the supernova envelope (Reynolds \& Chevalier 1984; 
Kennel \& Coroniti 1984). The bubble pressure sweeps the envelope matter into the thin dense shell 
(TDS). The bulk of the observed continuum is presumbaly formed in this TDS. The model may be tested by 
comparing TDS radius to the photometric radius of the iPTF14hls photosphere reported by Arcavi et al. 
(2017), while the velocity of the unperturbed supernova gas at the boundary of the TDS 
should be consistent with the minimal gas velocity recovered from the H$\alpha$ profile.

The gas ionization and heating is brought about by relativistic particles penetrating into 
the envelope from the bubble. It is noteworthy that the bubble 
energy loss via synchrotron radiation unlikely could be 
dominant since this mechanism requires too large energy of relativistic electrons. Indeed, 
the synchrotron cooling time is $t_{syn} = 36E_{14}^{-1}B_{-3}^{-2}$ days (where $E_{14}$ is the 
electron energy in $10^{14}$ eV, $B_{-3}$ is the magnetic field strength in $10^{-3}$ Gauss). 
Below we omit details of the energy deposition in the envelope; instead we describe this process in terms of the loss time for the bubble energy, assuming the lost energy is 
deposited into the envelope. 

%---------------------------------------------------------------

\begin{table}[t]

\vspace{6mm}
\centering
{{\bf Table 1.} Parameters of light curve models}
\label{t-par} 

\vspace{5mm}\begin{tabular}{l|c|c|c|c|c} \hline\hline
Model  &   $M$         & $E$            &  $R_0$        & $ L_j$           &  $t_b$ \\
        &  $M_{\odot}$  & $10^{51}$ erg  & $10^{13}$ cm  & $10^{42}$ erg/s & days    \\
 \hline
 mod100   &  30           &  8             &  9            & 9.6           &    14  \\
 mod450   &  30           &  8             &  2            & 9             &   11  \\
  \hline
\end{tabular}
\end{table}
%---------------------------------------------------------------

%xxxxxxxxxxxxxxxxxxxxxxxxxxxxxxxxxxxxxxxxxxxxxxxxxxxxxxxxxxxxxx
\begin{figure}[h]
\centering
\hspace{-1cm}
\includegraphics[width=14cm]{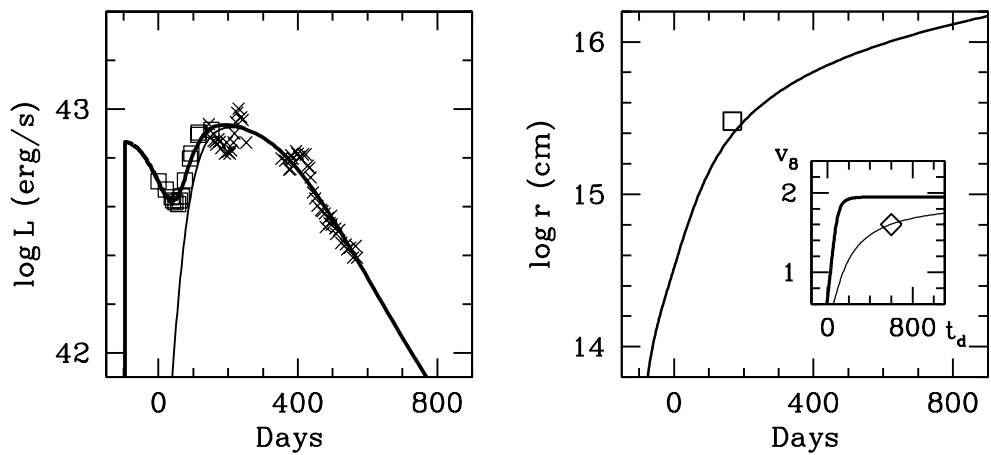}
\vspace{-3cm}
\caption{\footnotesize 
	Bolometric light curve and evolution of the bubble boundary in the
	mod100 model. {\it Left}: model light curve ({\it thick}) with two 
	luminosity regimes. The initial peak is due to the light from the explosion.
	The major luminosity maximum is powered by the injection of RBO into the envelope 
	({\it thin} line);
	 {\it crosses} are the data on the bolometric luminosity (Arcavi et al. 2017);
	{\it squares} are the R light curve (Arcavi et al. 2017) that is matched to the 
	bolometric luminosity around day 150. {\it Right}: evolution of the thin shell radius;
	square is the average of several estimates of the photosphere radius around day 167 
	(Arcavi et al. 2018). {\it Inset}: velocity evolution (ordinate in $10^8$ cm/s) 
	of the thin shell ({\it thick} line) and the mininimal velocity of the undisturbed 
	supernova gas. {\it Diamond} is the minimal velocity recovered from H$\alpha$ 
	on day 600.
	}
	\label{f-lc1}
\end{figure}
%========================================================
%xxxxxxxxxxxxxxxxxxxxxxxxxxxxxxxxxxxxxxxxxxxxxxxxxxxxxxxxxxxxxx

%xxxxxxxxxxxxxxxxxxxxxxxxxxxxxxxxxxxxxxxxxxxxxxxxxxxxxxxxxxxxxx
\begin{figure}[h]
\centering
\hspace{-1cm}
\includegraphics[width=14cm]{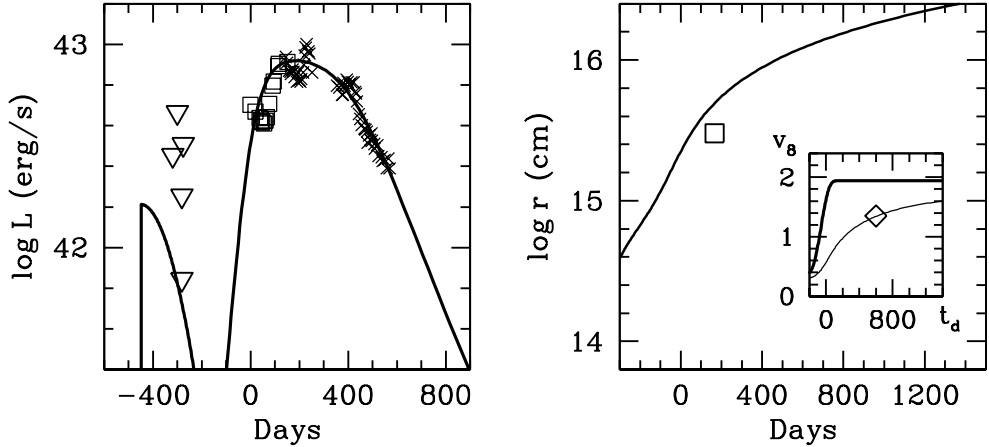}
\vspace{-3cm}
	\caption{\footnotesize
	The same as Figure 1, but for the mod450 model. In the {\it left} figure 
	{\it triangles} show upper limits of luminosity $\nu L_{\nu}$ recovered from 
	upper limits in R band.  
	}
	\label{f-lc1}
\end{figure}
%========================================================
%xxxxxxxxxxxxxxxxxxxxxxxxxxxxxxxxxxxxxxxxxxxxxxxxxxxxxxxxxxxxxx

\subsection{Light curve}

We consider two possible versions for the explosion moment: 100 days before the 
detection (mod100 model) and 450 days before the detection (mod450 model).
The light curve related to the explosion is calculated using the analytical 
solution (Arnett 1980) that is specified by the mass ($M$), the energy ($E$) and 
the initial radius ($R_0$). The radiation diffusion stage lasts about 100 days. 
With some delay after the explosion the major stage ensues for the power injection 
by the RBO; the relativistic bubble forms with the energy $E_b$ and 
radius $r_b$. The bubble pressure sweeps the supernova gas into the TDS of the 
mass $M_s$ with the expansion rate described by the equation 
\begin{equation}
M_s\frac{dv_b}{dt} = 4\pi r_b^2 \left[p_b - \rho\left(v_b - 
\frac{r}{t}\right)^2\right]\,,
\label{eq-motion}
\end{equation}
with the relativistic pressure $p_b = E_b/4\pi r_b^3$ and ram pressure of supernova gas 
in the right-hand side.
The density vs. velocity distribution in the freely expanding supernova is 
set by the expression $\rho = \rho_0/[1+(v/v_0)^8]$ that approximates 
hydrodynamic modelling of SN~IIP (cf. Utrobin 2007). Parameters ($\rho_0$ and $v_0$) are 
specified by the mass and the kinetic energy. 

The energy equation for the bubble is 
\begin{equation}
\frac{dE_b}{dt} = L_j - E_b\frac{v_b}{r_b} - \frac{E_b}{t_b}\,.
\label{eq-energy}
\end{equation}
The first term in the right-hand side is the kinetic luminosity of RBO, the second is 
the pressure work, and the third is the energy loss due to the escape of relativistic
protons from the bubble into the envelope with a minor contribution of synchrotron losses
of electrons. 
We assume that the power deposited in the envelope is released as the bolometric 
luminosity $L = E_b/t_b$, which suggests that we ignore the radiation dynamic effect 
in the envelope. This approximation is justified at the late stages $>100$ days 
when the diffusion time gets smaller than the expansion age.
The time dependence of the RBO luminosity is assumed to be 
$L_j = L_0(t/t_1)^{k_1})/(1 + (t/t_1)^{k_1})(1 + (t/t_2)^{k_2})$ that describes 
an initial rise and a final drop of the RBO luminosity ($t_1$, $t_2$, $k_1$, and $k_2$ 
are free parameters). 

While searching for the optimal parameters (Table 1, Figures 1 and 2) we intend to 
minimize the envelope mass and the explosion energy. 
In line with the above mass lower limit we adopt $M = 30\,M_{\odot}$. 
The kinetic energy then is constrained by the H$\alpha$ 
absorption at the high velocities, which results in   
$E = 8\times10^{51}$ erg. The initial radius in the mod450 model cannot exceed 
significantly the value $R_0 = 2\times10^{13}$, otherwise the early luminosity powered by the explosion would be in conflict with the archive flux upper limits (Figure 2). 
By the same reason the higher mass and energy are less favoured. 
The characteristic time of the bubble energy loss is $t_b = 14$ days and 11 days in 
the mod100 and mod450 model, respectively, so as the minimal velocity of the 
unperturbed supernova gas to be consistent with the value $v_{min} = 1340$ km s$^{-1}$
implied by the H$\alpha$ (see below).
Remarkably, the radius of the TDS, presumably the major continuum source,  
is close to the photosphere radius estimated by Arcavi et al. (2017). The 
factor 1.5 larger radius in the mod450 model might stem from the model simplicity and 
uncertainty of estimates of the photosphere radius. Moreover, the TDS might experience 
a fragmentation due to the Rayleigh-Taylor instability at the acceleration stage 
thus resulting in the reduction of the effective photosphere radius.  
To summarize, both versions of the explosion moment turn out viable as yet.

%---------------------------------------------------------------

\begin{table}[t]

\vspace{6mm}
\centering
{{\bf Table 2.} Parameters of spectral models}
\label{t-par} 

\vspace{5mm}\begin{tabular}{l|c|c|c|c|c|c} \hline\hline
Model  & $T$    & $\tau_T$  &  $y$(Na\,I) &  $L(\mbox{H}\alpha)$  &  $x$ & 
$L(\mbox{[OI]})$  \\
        &  K     &           &  $10^{-4}$  &  $10^{41}$ erg/s      &      &  
 $10^{40}$ erg/s   \\
 \hline
mod100    &  5400  &  0.8    &  0.8          &   2.2              &   0.27 & 2.8\\
mod450    &  5700  &  0.8    &  1.5          &   3.1              &   0.6  & 4  \\
\hline
\end{tabular}
\end{table}
%---------------------------------------------------------------

%xxxxxxxxxxxxxxxxxxxxxxxxxxxxxxxxxxxxxxxxxxxxxxxxxxxxxxxxxxxxxx
\begin{figure}[h]
\centering
\hspace{-1cm}
\includegraphics[width=12cm]{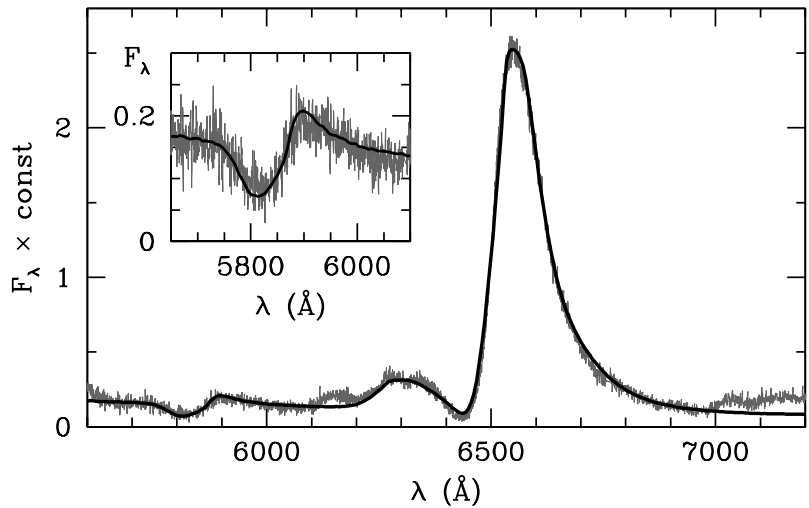}
\vspace{-3cm}
	\caption{\footnotesize
       iPTF14hls spectrum on day 600. Synthetic spectrum in the mod100 model 
	({\it thick} line) is compared to the observed spectrum (Arcavi et al. 2027).
	{\it Inset} shows the blowup of Na\,I doublet.
	}
	\label{f-lc1}
\end{figure}
%========================================================
%xxxxxxxxxxxxxxxxxxxxxxxxxxxxxxxxxxxxxxxxxxxxxxxxxxxxxxxxxxxxxx

%xxxxxxxxxxxxxxxxxxxxxxxxxxxxxxxxxxxxxxxxxxxxxxxxxxxxxxxxxxxxxx
\begin{figure}[h]
\centering
\hspace{-1cm}
\includegraphics[width=12cm]{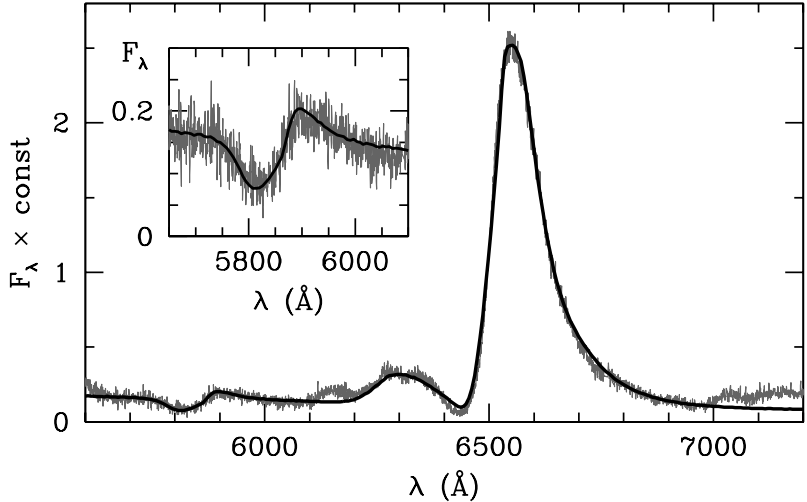}
\vspace{-3cm}
	\caption{\footnotesize
	The same as Figure 3, but for the mod450 model.
	}
	\label{f-lc1}
\end{figure}
%========================================================
%xxxxxxxxxxxxxxxxxxxxxxxxxxxxxxxxxxxxxxxxxxxxxxxxxxxxxxxxxxxxxx

\subsection{H$\alpha$ modelling}

We concentrate at the spectrum of iPTF14hls on day 600, the last in 
the presented set (Arcavi et al. 2017). The spectrum obtained at Keck 2 
is taken from the data base {\em http://wiserep.weizmann.ac.il} 
(Yaron \& Gal-Yam 2012). This moment corresponds to the age 700 and 1050 days 
for the models mod100 and mod450, respectively. At that late stage the resonance 
H$\alpha$ quanta scattering is essentially conservative, i.e., the quanta escape the 
local scattering without loss. The H$\alpha$ emission is dominated by 
the recombination, yet some contribution may come from the collisional excitation 
of $n > 2$ levels by the secondary and energetic photoelectrons. The ratio  
of the collisional excitation rate of these levels to the collisional ioniation 
is 0.14/0.39 = 0.36 (Xu et al. 1992). This suggests that in the recombination 
case C (optically thick Balmer lines) the total rate of the H$\alpha$ emission is by the factor 1.36 is larger than the recombination emission rate which is taken into account. 
Since the detailed ionization and heating processes are not considered here we 
assume a simple approximation of the constant ionization fraction over the envelope.
Formally this approximation corresponds to the requirement that the ionization rate 
per one hydrogen atom depends on the radius (velocity) 
as the density, $\zeta \propto \rho(v)$. 

The H$\alpha$ recombination coefficient in the case C (Osterbrock 1989)
is $\alpha_{32} = 3.1\times10^{-13}(T/5500)^{-0.89}$~cm$^3$\,s$^{-1}$,
 while the H$\alpha$ optical depth ($\tau_{23}$) is calculated assuming Boltzmann excitation with the temperature of 5400\,K and 5700\,K in models mod100 and mod450, respectively.
Noteworthy, in the inner zone, where $\tau_{23} > 1$, the actual behavior of the 
optical depth $\tau_{23}(v)$ does not affect the H$\alpha$ profile because 
of the absorption saturation and the conservative scattering in the line.

The continuum source in the model is associated with the TDS located at the  
minimal velocity of unperturbed supernova envelope $v_{min}$. The region $v < v_{min}$ is 
swept up into the TDS that is presumably partially fragmented and 
assumed to be semi-transparent with the optical depth $\tau_c = 0.5$. 
The H$\alpha$ photons are 
emitted only by the gas with velocities $v > v_{min}$. Apart from continuum and H$\alpha$ we include the Na\,I doublet with the fraction of neutrals $y$(Na\,I) 
assumed to be constant over the envelope (Table 2).
We include also oxygen doublet [O\,I] 6300, 6364 \AA. Its emissivity is assumed to be 
proportional to the density, while the integral doublet luminosity is set as some fraction of H$\alpha$ 
luminosity. The [O\,I] doublet luminosity in Table 2 is not corrected for the 
blending effect of Fe\,II lines (cf. Section 3).

Monte Carlo modelling (Figures 3, 4, Table 2) assuming the envelope of 
the light curve model shows good agreement with the observed spectrum on day 600 
for $v_{min} = 1350$ km s$^{-1}$ and $\tau_T =0.8$. The H$\alpha$ luminosity is 
$2.2\times10^{41}$ erg s$^{-1}$ and  $3.1\times10^{41}$ erg s$^{-1}$, while the ionization fraction 
is 0.27 and 0.6 for the mod100 and mod450 models, respectvely. Note that the H$\alpha$ luminosity 
in the mod450 model coincides with that of Dessart (2018), which in turn describes 
absolute fluxes on day 600. 
In the mod100 model the H$\alpha$ luminosity is 1.5 times lower, which 
makes the mod450 preferred. However given an approximate character of models the 
mod100 should not be finaly discarded.
It is remarkable that the Na\,I model profile fit the observational 
profile fairly well including velocity of the absorption minimum 
(4000 km s$^{-1}$) despite 
the low velocity of the continuum source (1350 km s$^{-1}$). The required 
fraction of neutral Na is larger than that calculated from Saha for $T = 5500$\,K by a factor of ten, the expected over-recombination effect 
for sodium due to the low photoionization cross section from the Na\,I ground level.

\section{Synthesised oxygen}

The above estimate of the oxygen doublet luminosity ignores blending by 
Fe\,II (6150, 6175, 6229 É 6239 \AA) lines, that are apparent in the blue part of the doublet.
In Figure 5 two versions of the doublet decomposition are shown for the model mod450:
assuming opticaly thin and optically thick doublet lines. The contribution of Fe\,II lines 
is described using Gaussians with the Doppler broadening $b = 1300$ km s$^{-1}$ 
and the blue shift 
of -280 km s$^{-1}$. In both cases doublet components are represented by shifted Gaussians 
(Table 3). 
In the optically thin case the doublet ratio is $I(6300)/I(6364) = 2.93$ and each line
of the [O\,I] doublet is described by two components: blue ($v_s = -1100$ km s$^{-1}$, 
$ b = 1500$ km s$^{-1}$) and red ($v_s = +1750$ km s$^{-1}$, $ b = 1700$ km s$^{-1}$).
In the optically thick case the doublet ratio is $I(6300)/I(6364) = 1.2$ (for $T =5700$\,K) 
one blue shifted component is sufficient ($v_s = -900$ km s$^{-1}$, $ b = 1650$ km s$^{-1}$).
The doublet luminosity is both cases is $2.2\times10^{40}$ erg s$^{-1}$. For the 
kinetic temperature of 5700\,K this corresponds to the neutral oxygen mass of
$1.2\,M_{\odot}$. Adopting the same ionization fraction as for hydrogen (i.e. 0.6) 
one gets the total oxygen mass of $3\,M_{\odot}$.

Parameters of the [O\,I] emissivity distribution (Table 3) suggest that the bulk of the 
oxygen resides in the central zone with the radius of about 2000 km s$^{-1}$.
For the homogeneous distribution of $1.2\,M_{\odot}$ of the neutral oxygen 
the local optical depth in the [O\,I] 6300 \AA\ is then 0.01. 
The optically thick option with the ratio $I(6300)/I(6364) = 1.2$ is realized 
for the clumpy distribution with the filling factor $f = 2\times10^{-3}$, in 
which case the ionization is low, so the oxygen is mostly neutral.
Thus, depending on the filling factor the oxygen mass estimate turns out to be in the 
range $1-3\,M_{\odot}$ for the kinetic temperature of 5700\,K.

The host galaxy metalicity $(0.4 - 0.9)\mbox{Z}_{\odot}$ (Arcavi et al. 2017) suggests 
that the initial oxygen mass in the $30\,M_{\odot}$ hydrogen envelope must be in the 
range $(0.15 - 0.3)\,M_{\odot}$, one order of magnitude lower than the above estimate.
This implies that in the iPTF14hls we see the oxygen synthesised in the course of
the star evolution. One should note that the oxygen temperature may well be lower 
than the accepted value because of the efficient cooling by metals, so the 
obtained oxygen mass estimate should be considered rather as a lower limit.
Noteworthy, the ejected oxygen might be only a fraction of all the 
precollapsed oxygen mantle, significant part of which could collapse along with the core.

Remarkable fact is that regardless of the oxygen filling factor, the oxygen is 
distributed non-spherically. In the optically thin case the distribution is bimodal with the 
dominant blue component and strongly shifted weaker red component.
In the optically thick case a single blue shifted component can describe the 
doublet profile.
The oxygen asymmetry could be caused by either the explosion asymmetry, or by the 
dynamic effect of BPO at the major luminosity stage.
In principle, the blue shift of [O\,I] doublet might be related ti the 
dust formation in the oxygen-rich material likewise in SNe~IIP, e.g., 
SN~1987A and SN~1999em. However in the iPTF14hls at the epoch of interest the dust 
formation is forbidden because of high temperature, moreover, we do not see 
similar asymmetry in the H$\alpha$ profile.

%---------------------------------------------------------------

\begin{table}[t]

\vspace{6mm}
\centering
{{\bf Table 3.} Parameters of oxygen distribution}
\label{t-par} 

\vspace{5mm}\begin{tabular}{l|c|c|c} \hline\hline
Regime        & Component  & $v_s$    &  $b$\\
             &           &  km/s   &  km/s \\
 \hline
 $\tau \ll 1$   &  1      & -1100   &   1500  \\
                &  2      &  +1750  &   1700  \\
 $ \tau \gg 1$   &         & -900    &   1650  \\
\hline

\end{tabular}
\end{table}
%---------------------------------------------------------------
%xxxxxxxxxxxxxxxxxxxxxxxxxxxxxxxxxxxxxxxxxxxxxxxxxxxxxxxxxxxxxx
\begin{figure}[h]
\centering
\hspace{-1cm}
\includegraphics[width=12cm]{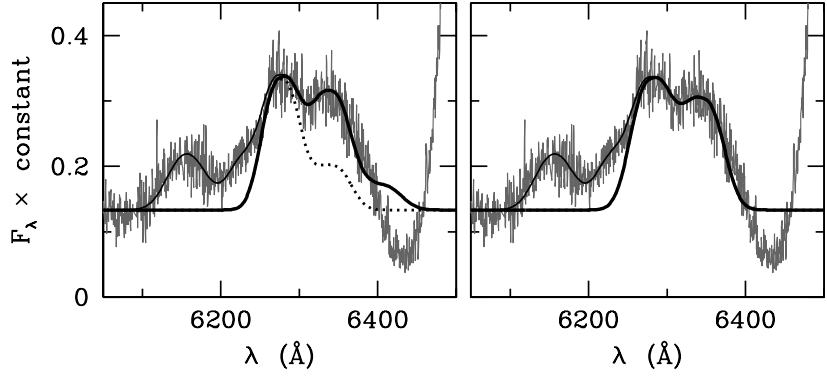}
\vspace{-3cm}
	\caption{\footnotesize	
	[O\,I] 6300, 6364 \AA\ doublet on day 600. {\it  Left}: the case 
	of optically thin lines; {\it dotted} line shows doublet for the blue 
	component. {\it Right}: opticaly thick case. In both cases {\it thin} line 
	shows the Fe\,II lines contribution.
	}
	\label{f-lc1}
\end{figure}
%========================================================

\section{Discussion and Conclusions}

The aim of the paper has been to find from the available observational data 
on the iPTF14hls (Arcavi et al. 2017) additional arguments in favour of either of two 
proposed scenarios: (i) the ejection of a massive envelope of a very massive star 
several hundred days before the discovery with the subsequent power deposition 
into the envelope from the black hole accretion (Arcavi et al. 2017); 
(ii) the explosion of moderately massive star coeval with the discovery  
accompanied by the ejection of the $13~M_{\odot}$ envelope powered by a magnetar 
(Dessart 2018). We demonstrated that the first scenario with the explosion 
around 450 days before the detection is preferred although the explosion 
100 days before the discovery cannot be rulled out as yet. As optimal, we adopt  
the explosion of a massive star with the radius of $2\times10^{13}$~cm 450 days before 
the discovery and the ejection of $30\,M_{\odot}$ envelope with 
the kinetic energy of $8\times10^{51}$ erg. Estimates of the mass and energy may be 
uncertain within 30\% (with the constant $E/M$ ratio). 

Following Arcavi et al. (2017) we adopt scenario with the black hole accretion 
to be the source of power for the major luminosity stage. As a next step, we 
suggest that the deposition of the accretion power into the envelope occurs in the form 
of the RBO which create the relativistic bubble in the central zone; relativistic 
protons are injected then into the envelope providing its ionization and heating. 

We see in [O\,I] 6300, 6364 \AA\ lines the emission of at least $1-3\,M_{\odot}$ 
of synthesised oxygen ejected by the explosion. Its distribution in the inner zone 
($v \lesssim 2000$ km s$^{-1}$) is essentially non-spherical with two componets 
in the optically thin case and one blue component in the optically thick case.
The asymmetry of the oxygen-rich gas could be caused by either explosion or the 
dynamic effect of
the BPO. Interestingly, that a similar asymmetry is seen in the [O\,I] doublet 
in the late time spectrum on day 1153 (Andrews \& Smith 2018). It seems plausible 
that at this late stage 
we observe the emission of the same oxygen-rich gas from the inner envelope  
powered by the black hole accretion mechanism. Yet some contribution in the heating on day 1153 (age of 1603 d) is related to the shock wave driven by the TDS with the kinetic luminosity of $2\times10^{39}$ erg s$^{-1}$ in the mod450 model. 
Boxy H$\alpha$ emission in the spectrum on day 1153  with the velocity range 
between -2000 and +1700 km s$^{-1}$ may well be emitted by the fragmented TDS with the
expansion velocity of 1900 km s$^{-1}$ in the mod450 model (Figure 2).

Despite the absence of close analogues among known supernovae, one should note that 
the two specific features of iPTF14hls -- massive ($\sim30\,M_{\odot}$) hydrogen-rich 
ejecta and high kinetic energy ($\sim10^{52}$ erg) --  is reminiscent of peculiar type 
IIP supernova SN~2009kf for which an explosion due to a rapid disk accretion onto 
the black hole (by analogy with hypernovae) has been invoked (Utrobin et al. 2010). 
Possibly the same mechanism is responsible for the explosion and ejection of massive 
envelope of iPTF14hls. However at this point resemblance comes to an end.  
Why the second stage of the disk accretion resposible for powerful long-lived
luminosity was turned on remains enigmatic.

\vspace{1cm}
I am grateful to Victor Utrobin and Sergei Blinnikov for discussions, and to 
Luc Dessart for kindly sending synthetic spectrum. For this work 
"Weizmann interactive supernova data repository" has been very usefull.

\pagebreak   
%****************************************************************

%==================================================================================

\end{document}